\documentclass[a4paper]{article}

\usepackage{INTERSPEECH2020}
\usepackage{xspace}
\usepackage{lipsum}
\usepackage{subcaption}
\usepackage{siunitx}
\usepackage{hyperref}
\usepackage{multirow}
\usepackage{pbox}
\usepackage{tabularx}
\usepackage{xcolor}

\title{%
	Cotatron:
	Transcription-Guided Speech Encoder \\
	for Any-to-Many	Voice Conversion without Parallel Data
}
\name{Seung-won Park$^{1, 2}$, Doo-young Kim$^{1, 2}$, Myun-chul Joe$^{2} $}
\address{
	$^1$Seoul National University \qquad $^2$MINDsLab Inc.
}
\email{\{swpark, dykim, mcjoe\}@mindslab.ai}

\makeatletter
\DeclareRobustCommand\onedot{\futurelet\@let@token\@onedot}
\def\@onedot{\ifx\@let@token.\else.\null\fi\xspace}

\def\eg{\emph{e.g}\onedot} 
\def\ie{\emph{i.e}\onedot}

\def\etal{\emph{et al}\onedot}
\makeatother

\newcommand{\mel}{mel spectrogram\xspace}
\begin{document}

\maketitle
\begin{abstract}
	We propose \textit{Cotatron},
	a transcription-guided speech encoder for
	speaker-independent linguistic representation.
	Cotatron is based on the multispeaker TTS architecture
	and can be trained with conventional TTS datasets.
	We train a voice conversion system
	to reconstruct speech with Cotatron features,
	which is similar to the previous methods based on Phonetic Posteriorgram (PPG).
	By training and evaluating our system with 108 speakers from the VCTK dataset,
	we outperform the previous method in terms of
	both naturalness and speaker similarity.
	Our system can also convert speech from 
	speakers that are unseen during training,
	and utilize ASR to automate the transcription
	with minimal reduction of the performance.
	Audio samples are available at \url{https://mindslab-ai.github.io/cotatron},
	and the code with a pre-trained model will be made available soon.
\end{abstract}

\section{Introduction}

Recent advances in voice conversion (VC) have shown
potential for a wide variety of applications,
such as enhancement of impaired speech or entertainment purposes.
To switch the source speech's speaker identity to that of the target speaker,
the system should be able to encode speaker-independent
(\eg, linguistic)
features from given speech,
and then pair them with a speaker representation to reconstruct the speech.
Phonetic Posteriorgram (PPG) \cite{ppg},
a speaker-independent feature extracted with speaker-independent ASR,
had been widely used for non-parallel voice conversion
\cite{saito2018non, yeh2018rhythm, lian2019towards, lu2019one}.
However, PPG-based methods
usually required additional acoustic features from audio analysis,
which may indicate that the PPG itself is insufficient
to encode rich linguistic features of human speech.

One way to
encode speaker-independent features without discarding essential factors of the speech
is to train a speech encoder with some restrictions.
For example, Qian \etal \cite{autovc} showed that 
an autoencoder with a carefully tuned bottleneck
can effectively encode speaker-independent features
without losing content information.
Other prior works on restricting the speech encoders include:
propagating reversed gradient from the speaker classifier \cite{chou2018multi},
applying instance normalization \cite{chou2019one},
quantizing the representation \cite{vqvae, ding2019group, liu2019unsupervised},
and training a conditioned flow-based generative model \cite{blow}.
However, these methods required the model to discover linguistic representations by itself,
although transcriptions were available within the dataset.

Initial attempts on incorporating text supervision to voice conversion system
trained an auxiliary ASR decoder \cite{zhang2019improving, zhang2019non, parrotron},
or shared the model weights with TTS \cite{zhang2019joint}.
Unfortunately,
these methods only dealt with a limited number of speakers
or required huge amounts of data for each speaker;
making their effectiveness on the real-world applications questionable.

In this paper, we propose \textit{Cotatron},
a transcription-guided speech encoder
based on a pre-trained multispeaker TTS model \cite{tacotron2, sv2tts}.
Cotatron encodes an arbitrary speaker's speech
into speaker-independent linguistic features,
which are fed to a decoder for non-parallel, any-to-many voice conversion.
Our Cotatron-based voice conversion system outperforms the previous state-of-the-art method,
Blow \cite{blow},
in terms of both naturalness and speaker similarity scores on a user study,
when trained and evaluated with 108 speakers from VCTK dataset \cite{vctk}.

\section{Approach} \label{sec:model}

\subsection{Speaker-independent linguistic features from TTS} \label{subsec:bmm}

\begin{figure}[t]
	\centering
	\includegraphics[width=\linewidth]{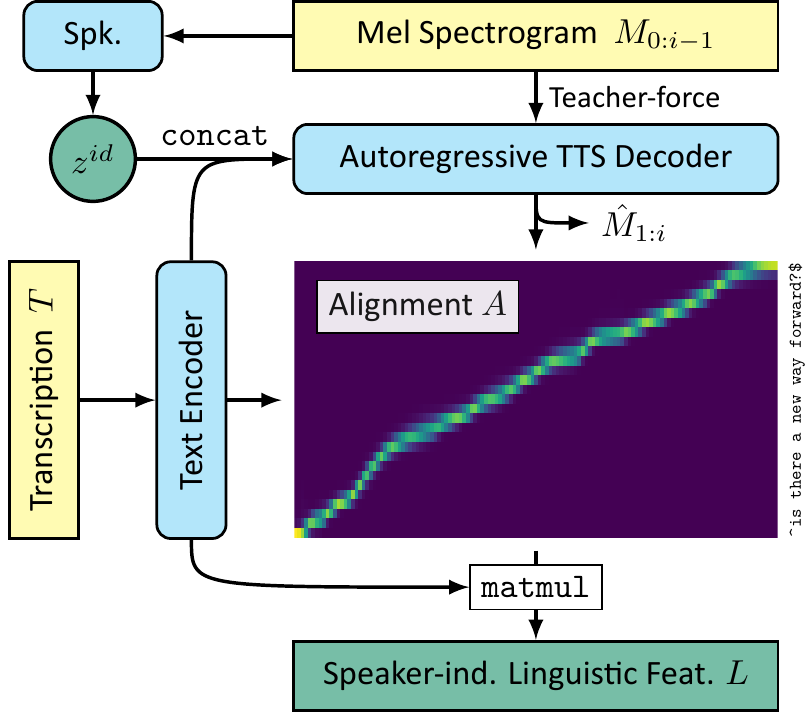}
	\caption{
		Cotatron architecture.
		The alignment between \mel and its transcription
		is obtained via a pre-trained multispeaker TTS (Tacotron2)
		and then combined with text encoding
		to extract speaker-independent linguistic features.
		\emph{Spk.} denotes the speaker encoder.
	}
	\label{fig:linguistic}
\end{figure}

Cotatron is guided with a transcription
to extract speaker-independent linguistic features from the speech.
Cotatron's basic architecture is identical to multispeaker Tacotron2 \cite{tacotron2, sv2tts};
it jointly learns to align and predict next mel frame
from the text encoding, previous mel frame, and the speaker representation:
\begin{equation}\label{eq:tacotron2}
	\hat{M}_{1:i}, A_{i} = \mathrm{\underset{tts}{Decoder}}\left(
		\mathrm{\underset{text}{Encoder}}\left(T\right),
		M_{0:i-1}, z^{id}\right),
\end{equation}
where $ T, M, A, z^{id} $ corresponds to text,
log \mel, alignment, and the speaker representation, respectively.

After training, a simple yet effective trick is applied.
An alignment $ A $ between the speech and the transcription is obtained
via feeding all frames of the \mel into Cotatron with teacher-forcing applied.
Then, the speaker-independent linguistic features of the speech are obtained from
a matrix multiplication of
the alignment and text encoding as Fig. \ref{fig:linguistic}:
\begin{equation}\label{eq:bmm}
	L = \mathtt{matmul}\left(A, \mathrm{\underset{text}{Encoder}}(T)\right).
\end{equation}
Per definition, the text encoding contains no speaker information.
Besides, the text-audio alignment $ A $
is a set of scalar coefficients for weighted summation over encoder timesteps of the text encoding.
Hence, we may argue that the Cotatron features $ L $
do not explicitly contain a source speaker's information.
We show the degree of speaker disentanglement at Sec. \ref{subsec:disentanglement}.

Cotatron features are naturally adequate for
synthesizing speech from a large number of speakers;
the features can be interpreted as
context vectors for Tacotron2's attention mechanism,
which are already optimized for multispeaker speech synthesis.
We further expand the coverage of source speakers into arbitrary
by replacing
the embedding table into an encoder for speaker representation $ z^{id} $.
The speaker encoder is composed of 6 layers of 2D CNN,
following the reference encoder architecture from
Skerry-Ryan \etal \cite{e2eprosody}.
Each layer had $ 3\times3 $ kernel, $ 2\times2 $ stride
with 32, 32, 64, 64, 128, 128 channels.
The CNN output is flattened and
passed through a 256-unit GRU
to obtain the fixed-length speaker representation
from the final state.

\subsection{Voice conversion}

\subsubsection{Residual Encoder}
Let's consider a decoder reconstructing the speech from Cotatron features.
Even when the rhythm of the transcription is given via Cotatron features,
other components of the speech may vary.
For example, the intonation may vary
within the speech of the same text with identical rhythm.
It is therefore insufficient for the decoder 
to use only the Cotatron features and the speaker representation.
To fill the gap of information,
we design an encoder to provide decoder a residual feature $ R $.

The residual encoder (Fig. \ref{fig:residual_encoder}) is built with 6 layers of 2D CNN
as the speaker encoder did,
but strides are not applied across time
to preserve the temporal dimension of the \mel.
Each layer had $ 3\times3 $ kernel with $ 2\times1 $ stride
and 32, 32, 64, 64, 128, 128 channels.
If the dimension of the residual features is too wide,
the residual encoder may learn to cheat by encoding the information
that is related to the individual speaker -- \eg, absolute pitch.
We find that a single-channeled output helps
to prevent the residual features from containing
characteristic of the individual speaker,
and is enough to represent residual information of the speech;
this approach was also used by Lian \etal \cite{lian2019towards}.
After projecting to a single channel,
instance normalization \cite{inorm} is applied
to prevent the residual representation from containing speaker-dependent information.
Finally,
the values are smoothed after tanh activation by applying convolution with a Hann function
of window size 21.

\subsubsection{VC Decoder} \label{subsubsec:decoder}

\begin{figure}[t]
	\centering
	\begin{subfigure}[t]{\linewidth}
		\centering
		\includegraphics[width=\linewidth]{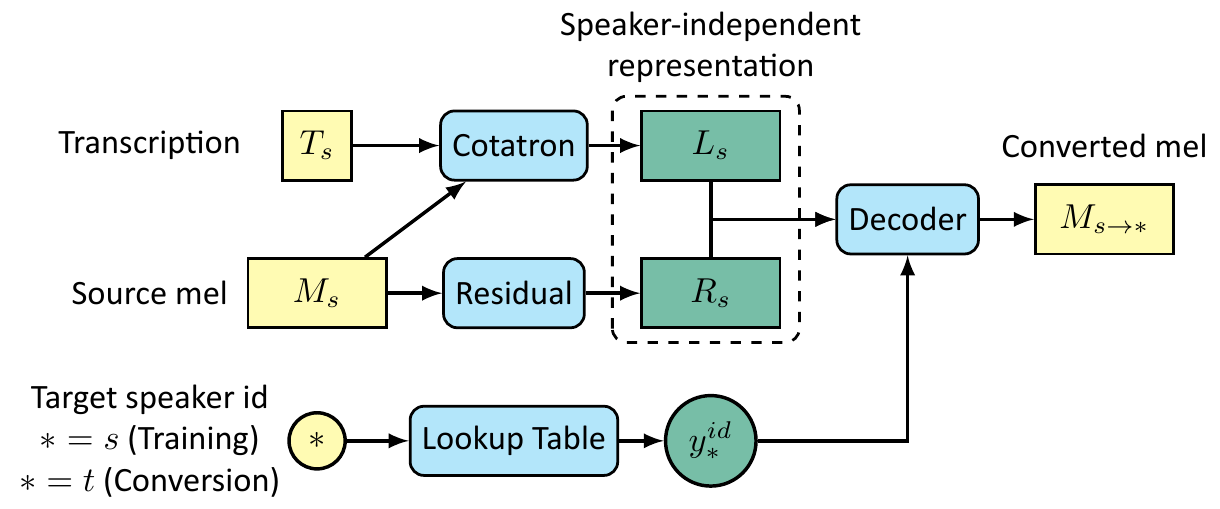}
		\caption{Voice Conversion system with Cotatron.}
		\label{fig:overall}
	\end{subfigure}
	\vspace{0.2cm}
	
	\hspace{-0.4cm}
	\begin{subfigure}[t]{0.49\linewidth}
		\centering
		\includegraphics[height=0.21\textheight]{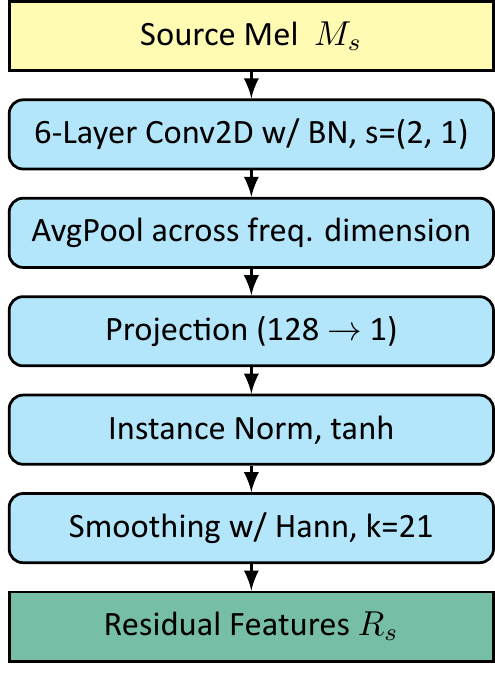}
		\caption{Residual encoder.}
		\label{fig:residual_encoder}
	\end{subfigure}
	\begin{subfigure}[t]{0.49\linewidth}
		\centering
		\includegraphics[height=0.21\textheight]{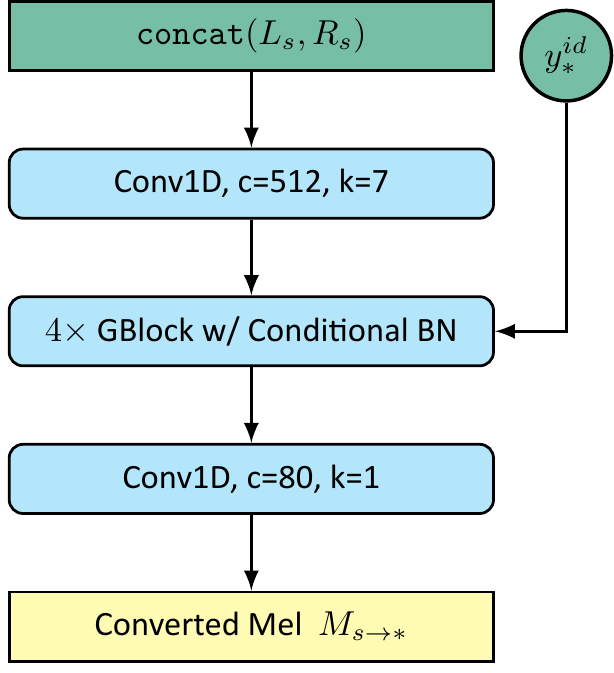}
		\caption{VC decoder.}
		\label{fig:decoder}
	\end{subfigure}
	\caption{%
		Network architectures.
		$ s, c, k $ denotes the
		stride, number of channels, kernel size
		of convolution layer, respectively.
		Speaker representation $ y^{id} $ conditions the VC decoder
		via conditional batch normalization layer
		within residual blocks.
		Refer to Binkowski \etal \cite{gantts}
		for the detailed architecture of GBlock.
	}
	\label{fig:mel_encoder}
\end{figure}

The decoder for voice conversion (Fig. \ref{fig:decoder}) is trained to reconstruct the \mel
from a given pair of information;
the Cotatron features $ L $ 
and the residual feature $ R $ 
are concatenated channel-wise,
and then conditioned with
256-dimension speaker embedding $ y^{id} $ retrieved from a lookup table as:
\begin{equation}\label{eq:decoder}
	M_{s\rightarrow\ast} = \mathrm{\underset{vc}{Decoder}}\left( \mathtt{concat}\left(L_{s}, R_{s}\right), y_{\ast}^{id} \right).
\end{equation}
The asterisk symbol
can be either $ s $ or $ t $,
each representing source/target of the voice conversion.
Thus, $ M_{s\rightarrow s} $ denotes reconstruction,
and $ M_{s\rightarrow t} $ denotes voice conversion from $ s $ to $ t $.

Following the model architecture of GAN-TTS \cite{gantts},
the VC decoder is constructed with
a stack of four \textit{GBlock}s without upsampling.
Each GBlock has 512, 384, 256, 192 channels, respectively.
For speaker conditioning, the embedding of the target speaker $ y^{id} $
is injected via a conditional batch normalization layer \cite{condbn}
within the GBlocks,
after an affine transformation.
We empirically observed that concatenation of speaker embedding leads to worse results.
Neither the hyper-conditioning \cite{hypernet} nor the weight demodulation \cite{stylegan2} did not help.
There might be room for improvement in design choices of decoder architecture,
but we leave it as a future work since it is beyond the scope of this work.

Note that the VC decoder is only trained to reconstruct the \mel with representations from the identical speaker.
Though it is possible to directly train the conversion in an adversarial manner,
we show the effectiveness of Cotatron on voice conversion
using only reconstruction loss.

\section{Experimental setup}

\subsection{Dataset}
Our voice conversion system is trained and evaluated with a VCTK dataset \cite{vctk},
which consists of 46 hours of English speech from 108 speakers.
Similar to what Blow had done \cite{blow},
we split the data into train, validation, and test splits
by randomly selecting 80\%, 10\%, 10\% of the data, respectively.
To prevent overlap of transcription between data splits,
the data is split with respect to their transcription,
not the number of files.

To stabilize the training of multispeaker TTS,
we incorporate a subset of LibriTTS \cite{libritts},
which is a dataset specialized in training TTS systems.
Speakers with more than 5 minutes of speech
are chosen from LibriTTS' \textit{train-clean-100} subset.

All audios longer than 10 seconds are not used for training
to allow efficient batching.
The audios are resampled to sampling rate \SI{22.05}{\kilo\hertz}
and then normalized without silence removal.
The statistics of the dataset are shown in Table \ref{tab:dataset}.

\begin{table}[h]
	\caption{%
		Dataset statistics.
		For LibriTTS \emph{train-clean-100} split,
		speakers with less than 5 minutes of speech are removed.
	}
	\label{tab:dataset}
	\begin{tabularx}{\linewidth}{Xcc}
		\toprule
		\textbf{Dataset} & \textbf{\# speakers} & \textbf{Length (h)} \\
		\midrule
		VCTK \cite{vctk} train / val / test & 108 & 34.6 / 4.5 / 4.2 \\
		\midrule
		LibriTTS \cite{libritts} && \\
		\emph{train-clean-100} & 123 & 23.4 \\
		\emph{dev-clean} & \phantom{0}40 & \phantom{0}9.0 \\
		\emph{test-clean} & \phantom{0}39 & \phantom{0}8.6 \\
		\bottomrule
	\end{tabularx}
\end{table}

\subsection{Training}

\subsubsection{Cotatron}
Cotatron is trained with the aforementioned subset of LibriTTS,
which is based on the \emph{train-clean-100} split.
Then, the model is transferred to learn with both LibriTTS and VCTK train split.
To enhance the stability of text-audio alignment learning,
the autoregressive decoder is teacher-forced with a rate of 0.5,
\ie, input mel frame is randomly selected from
either ground truth frame or previously generated frame.
Furthermore, we find it helpful to train extra MLP with dropout
for speaker classification on top of $ z^{id} $ from the speaker encoder,
using cross-entropy loss $ \mathcal{L}_{id} $.
Overall, Cotatron is trained with
the sum of \mel reconstruction loss and speaker classification loss:
\begin{equation}\label{eq:cotatron_loss}
	\mathcal{L}_{\text{cotatron}} =
	\left\| \hat{M}_{s, pre} - M_{s} \right\|_{2}^{2}
	+ \left\| \hat{M}_{s, post} - M_{s} \right\|_{2}^{2}
	+ \mathcal{L}_{id},
\end{equation}
where $ \hat{M}_{s, pre} $ and $ \hat{M}_{s, post} $
denote output before and after
the Cotatron's post-net \cite{tacotron2}, respectively.

Throughout the training process, Adam optimizer \cite{adam} is used
with batch size 64.
The initial learning rate \num{3e-4} is used for the first 25k steps
and then exponentially decayed to \num{1.5e-5} for the next 25k steps.
After the model converges with LibriTTS, we add VCTK
and reuse the learning rate decay scheme.
Weight decay of \num{1e-6} is used for Adam optimizer,
and the gradient is clipped to 1.0 to prevent gradient explosion.

\subsubsection{Mel-Spectrogram Reconstruction}
After the training of Cotatron,
the components for voice conversion system is trained
on top of Cotatron features.
The residual encoder and the VC decoder is jointly trained
with \mel reconstruction loss:
\begin{equation}\label{eq:reconstruction}
	\mathcal{L}_{\text{vc}} = \left\| M_{s\rightarrow s} - M_{s} \right\|_{2}^{2}.
\end{equation}
During the reconstruction training phase,
Cotatron is set to evaluation mode;
all dropout layers are turned off,
and the autoregressive decoder is always teacher-forced
to provide consistent features for VC decoder.
Adam optimizer with constant learning rate \num{3e-4}
is used with weight decay \num{1e-6} and batch size 128.
Gradient clipping is not used here.

\subsection{Conversion}

To convert one voice to another,
we first extract the speaker-independent features, $ L_{s}, R_{s} $,
from the source speech with Cotatron and residual encoder, respectively.
Then, the embedding of the target speaker $ y_{t}^{id} $
is retrieved from the lookup table.
Finally, a pair of speaker-independent features and target speaker embedding
is used to produce a converted \mel, $ M_{s\rightarrow t} $.
The resulting \mel is then inverted into raw audio using MelGAN \cite{melgan},
which is trained with LibriTTS train split and then fine-tuned with the entire VCTK dataset.

\subsection{Implementation details}

For robust alignment stability against length variation,
we apply the Dynamic Convolution Attention (DCA) mechanism \cite{dca}.
The speaker representation is extracted from
the ground-truth \mel with the speaker encoder,
and then repeatedly concatenated with text encoder output
to feed the auto-regressive decoder of Cotatron.
For both Cotatron and the voice conversion system,
the training data is augmented with representation mixing \cite{kastner2019representation},
\ie, graphemes are randomly replaced with phonemes if the word is available in CMUdict \cite{cmudict}.
Both decoders produce 80-bin log \mel,
which is computed from \SI{22.05}{\kilo\hertz} raw audio using
STFT with window size 1024, hop size 256, Hann window,
and a mel filterbank spanning from \SI{70}{\hertz} to \SI{8000}{\hertz}.
The voice conversion systems are implemented with PyTorch \cite{pytorch}
and trained for 10 days with two NVIDIA V100 (32GB) GPU
using data parallelism.


\subsection{Evaluation metrics}

We validate the effectiveness of our method
with both subjective and objective metrics,
using 100 and 10,000 audio samples per each measurement, respectively.

\vspace{5pt}\noindent\textbf{Mean Opinion Score (MOS).}
To assess the naturalness of converted speech,
we measure the mean opinion score (MOS) on a 5-point scale at Amazon Mechanical Turk (MTurk).
A total of 100 audio samples are generated for each case
with a random pair of source speech and target speaker,
which contains all possible gender combinations.
The audio samples from our method and natural speech are downsampled to rate \SI{16}{\kilo\hertz}
to match the results from Blow \cite{blow}.
Each sample is assigned to 5 human listeners,
and the highest/lowest score is discarded.


\vspace{5pt}\noindent\textbf{Degradation Mean Opinion Score (DMOS).}
Another user study is done to assess speaker similarity
between the converted speech and the target speaker's original recording.
The degradation mean opinion score (DMOS) on a 5-point scale
is measured at MTurk with the same settings from the MOS experiment.

\vspace{5pt}\noindent\textbf{Speaker Classification Accuracy (SCA).}
Our system should be able to fool the
speaker classifier as if the converted speech
was spoken from the target speaker.
The speaker classifier is an
MFCC-based single-layer classifier,
which is identical to
the one used with Blow \cite{blow}
for a fair comparison.
The classifier is trained with
108 speakers from the VCTK train split
and achieved 99.4\% top-1 accuracy on the test split.
The MFCC is directly calculated from the log \mel if possible.

\vspace{5pt}\noindent\textbf{Voicing Decision Error (VDE).}
As a proxy metric for
content consistency between source and converted speech,
we measure the rate of voicing decision match between them,
adapting a metric of end-to-end prosody transfer for speech synthesis \cite{e2eprosody}.
The voicing decision is obtained via rVAD \cite{tan2020rvad}
with a VAD threshold value set to 0.7.

\section{Results}

\subsection{Many-to-many conversion}

We compare our system with Blow \cite{blow},
which is the only literature to date on many-to-many voice conversion
with all speakers of VCTK.
As presented in Table \ref{tab:evaluation},
our system shows significantly better results on
both MOS and DMOS than Blow,
even when only the Cotatron features are used without residual features.
Incorporating the residual encoder on our system has further enhanced the MOS.
It should, however, be noted that
the objective results on speaker similarity (SCA)
are contradicting that from the subjective results (DMOS).
Future work should, therefore, revisit and establish
objective speaker similarity metrics for voice conversion systems.

\begin{table}[h]
	\caption{
		Results of many-to-many voice conversion.
	}
	\label{tab:evaluation}
	\centering
	\begin{tabularx}{\linewidth}{lXccc}
		\toprule
		\multicolumn{2}{l}{\textbf{Approach}} & \textbf{MOS} & \textbf{DMOS} & \textbf{SCA} \\
		\midrule
		\multicolumn{2}{l}{Source as target}
		& $ 4.28 \pm 0.11 $ & $ 1.71 \pm 0.22 $  & \phantom{0}0.9\% \\
		\multicolumn{2}{l}{Target as target}
		& $ 4.28 \pm 0.11 $ & $ 4.78 \pm 0.08 $ & 99.4\% \\
		\midrule
		\multicolumn{2}{l}{Blow}
		& $ 2.41 \pm 0.14 $ & $ 1.95 \pm 0.16 $ & \textbf{86.8\%}  \\
		\multicolumn{2}{l}{\textbf{Cotatron} (ours)} &  &  & \\
		\hspace{-0.2cm} & w/o residual
		& $ 3.18 \pm 0.14 $ & $ \mathbf{4.06 \pm 0.17} $ & 73.3\% \\
		\hspace{-0.2cm} & \textbf{full model}
		& $ \mathbf{3.41 \pm 0.14} $ & $ 3.89 \pm 0.18 $ & 78.5\%   \\
		\bottomrule
	\end{tabularx}
\end{table}

\subsection{Any-to-many conversion and the use of ASR}

Considering the technical demands of the real-world applications,
we further explore the generalization power of our voice conversion system.
First, we consider any-to-many setting -- \ie,
converting arbitrary speakers' speech to that of speakers that are seen during training.
Next, we inspect the reliability of using ASR transcription,
which enables a fully automatic pipeline of our system without manual transcription.
For any-to-many conversion experiment,
we randomly sample speeches from LibriTTS \emph{test-clean} split
and convert them into speakers of VCTK.
For ASR, wav2letter++ \cite{zeghidour2018fully, wav2letter++} is used.

In Table \ref{tab:asr_unseen},
we present the MOS, SCA, and VDE for all possible cases of input.
First,
all of the MOS results are much better than the previous method in Table \ref{tab:evaluation},
though the scores from any-to-many setting are slightly lower than that of many-to-many setting.
Next, the differences of SCA across the cases are negligible,
and the values of VDE are minimal when considering the accuracy of the VAD module.
These results suggest that the conversion quality is rather unaffected by using
(1) source speech from speakers that are unseen during training,
and/or
(2) automated transcription from ASR.
Besides, it is surprising to observe that
the word errors of automated transcription
do not damage the performance;
this would seem to suggest that
most of the transcription errors originate from their homophones,
\eg, \textit{site} is often wrongly transcribed as \textit{sight}.

\begin{table}[h]
	\caption{
		Results of any-to-many conversion and using ASR transcription.
		The values are expected to be similar across the rows.
	}
	\label{tab:asr_unseen}
	\centering
	\begin{tabularx}{\linewidth}{Xccc}
		\toprule
		\textbf{Input Transcription} & \textbf{MOS} & \textbf{SCA} &  \textbf{VDE}  \\
		\midrule
		\multicolumn{3}{l}{\textit{VCTK test $ \rightarrow $ VCTK test (many-to-many)}} \\
		1-a. ground truth
		& $ 3.41 \pm 0.14 $ & 78.5\% & 2.98\% \\
		1-b. ASR (WER 12.6\%)
		& $ 3.44 \pm 0.12 $ & 77.8\% & 3.03\% \\
		\midrule
		\multicolumn{3}{l}{\textit{LibriTTS test-clean $ \rightarrow $ VCTK test (any-to-many)}} \\
		2-a. ground truth
		& $ 2.84 \pm 0.14 $ & 73.6\% & 11.9\% \\
		2-b. ASR (WER 7.0\%)
		& $ 2.83 \pm 0.15 $ & 71.7\% & 11.7\% \\
		 \bottomrule
	\end{tabularx}
\end{table}

\subsection{Degree of disentanglement} \label{subsec:disentanglement}
To quantify the degree of speaker disentanglement
of features from Cotatron and the residual encoder,
we additionally train a neural network
for classifying speakers from the VCTK dataset with a given set of features.
In the case of ideal speaker disentanglement,
the SCA will be close to that of random guessing: 0.9\%.
Each classification network is built with 4 layers of 1D CNN and batch normalization,
followed by the temporal max-pooling layer and MLP with dropout.

As shown in Table \ref{tab:disentanglement},
the SCA with Cotatron features and the residual features
are significantly lower than that from the source \mel.
These results indicate that our method effectively disentangles
the speaker's identity from the speech,
while it is noteworthy to mention that the network
was slightly able to guess the speaker using only Cotatron features.

\begin{table}[h]
	\caption{%
		Degree of speaker disentanglement.
	}
	\label{tab:disentanglement}
	\centering
	\begin{tabularx}{\linewidth}{Xcccc}
		\toprule
		\textbf{Input Feature} & Random  & $ L_{s} $ & $ (L_{s}, R_{s}) $ & $ M_{s} $ \\
		\midrule
		\textbf{SCA} & 0.9\% & 35.2\% & 54.0\% & 97.9\% \\
		\bottomrule
	\end{tabularx}
\end{table}

\section{Discussion}

In this paper, we proposed Cotatron,
a transcription-guided speech encoder for speaker-independent linguistic representation,
which is based on the multispeaker Tacotron2 architecture.
Our Cotatron-based voice conversion system
reaches state-of-the-art performance in terms of both naturalness and speaker similarity
on conversion across 108 speakers from the VCTK dataset
and shows promising results on conversion from arbitrary speakers.
Even when the automated transcription with errors is fed,
the performances remained the same.

To our best knowledge,
Cotatron is the first model
to encode the speaker-independent linguistic representation
by explicitly aligning the transcription with given speech.
This could open a new path towards multi-modal approaches for speech processing tasks,
where only the speech modality was usually being used.
For example, one may consider training
a transcription-guided speech enhancement system
based on Cotatron features.
Furthermore, traditional speech features that were utilized for lip motion synthesis
can be possibly replaced with Cotatron features
to incorporate the transcription for better quality.

Still, there is plenty of room for improvement in the voice conversion system with Cotatron.
Despite our careful design choices,
the residual encoder seems to provide speech features
that are entangled with speaker identity,
which may harm the conversion quality or even cause mispronunciation issues.
Besides, methods for conditioning the target speaker's representation
could be possibly changed;
\eg, utilizing a pre-trained speaker verification network as a speaker encoder
may enable any-to-any conversion with our system.

\section{Acknowledgments}

The authors would like to thank
Gaku Kotani from U. Tokyo,
June Young Yi, and Junhyeok Lee from MindsLab Inc.,
and other reviewers who elected to remain anonymous
for providing beneficial feedback on the initial version of this paper.

\bibliographystyle{IEEEtran}

\bibliography{swpark_is2020}

\end{document}